\documentclass[12pt]{article}

\usepackage{amssymb,amsmath}
\usepackage{graphicx}

\newcommand{\be}{\begin{equation}}
\newcommand{\ee}{\end{equation}}
\newcommand{\Dlt}{\Delta}

\newcommand{\prt}{\partial}

\newcommand{\bk}{{\bf k}}

\newcommand{\ba}{{\bf a}}

\newcommand{\bt}{\beta}

\newcommand{\ep}{\varepsilon}
\newcommand{\al}{\alpha}
\newcommand{\ra}{\rightarrow}
\newcommand{\sgm}{\sigma}

\newcommand{\om}{\omega}

\newcommand{\dgr}{\dagger}

\newcommand{\rgl}{\rangle}
\newcommand{\lgl}{\langle}

\begin{document}

\begin{center}
{\Large{\bf 
Quasiequilibrium Mixture of Itinerant and Localized Bose Atoms 
in Optical Lattice } \\ [5mm]

V.I. Yukalov$^a$, A. Rakhimov$^b$, and S. Mardonov$^c$} \\ [5mm]

{\it
$^a$ Bogolubov Laboratory of Theoretical Physics, \\
Joint Institute for Nuclear Research, Dubna 141980, 
Russia \\ [3mm]

$^b$ Institute of Nuclear Physics, Tashkent 100214, 
Uzbekistan \\[3mm]

$^c$ Samarkand State University, Samarkand, Uzbekistan}
\end{center}

\vskip 3cm

\begin{abstract}
Conditions are studied under which there can exist a quasiequilibrium 
mixture of itinerant and localized bosonic atoms in an optical lattice, 
even at zero temperature and at integer filling factor, when such a 
coexistence is impossible for an equilibrium lattice. The consideration 
is based on a model having the structure of a two-band, or two-component, 
boson Hubbard Hamiltonian. The minimal value for the ratio of on-site 
repulsion to tunneling parameter, necessary for the occurrence of such 
a mixture, is found.
\end{abstract}

\vskip 3cm

{\bf PACS numbers}: 03.75.Hh, 03.75.Kk, 03.75.Lm, 03.75.Mn, 
03.75.Nt, 05.30.Jp

\newpage

\section{Introduction}

Optical lattices provide exceptional opportunity for creating 
various states of periodic matter [1-5]. In this paper, we consider 
$N$ bosonic atoms in a lattice of $N_L$ sites, with the filling 
factor $\nu\equiv N/N_L$. The system is characterized by a boson 
Hubbard Hamiltonian, with on-site repulsion $U$, tunneling parameter 
$J$, and the number of nearest neighbors $z_0$. As is well known, 
Bose atoms at zero temperature and integer filling factor, form 
either Mott insulating state or superfluid state, depending on the 
ratio $u \equiv U/z_0 J$ of the on-site repulsion $U$ to the product 
of the tunneling parameter $J$ and the nearest-neighbor number 
$z_0$. For a cubic three-dimensional lattice, with $z_0 = 6$ and the 
unity filling factor ($\nu=1$), the second-order phase transition 
between superfluid and insulating states occurs at $u_c = 4.9$, as 
follows from strong-coupling perturbation theory [6,7] and Monte Carlo 
simulations [8-10]. At finite temperature and/or noninteger filling 
factor, there is the coexistence of localized and delocalized atoms 
[11-14].

Our concern in the present paper is to analyze a possible coexistence 
of delocalized, wandering, atoms and localized atoms in a lattice 
with an integer filling factor at zero temperature. As is known from 
the previous results, such a case cannot occur in an equilibrium lattice. 
Therefore, we need to keep in mind a kind of a nonequilibrium system. 

Let us assume that a nonequilibrium state has been prepared, where 
a portion of atoms is localized and another portion is not. This 
could be achieved in the process of loading atoms into the lattice. 
A nonequilibrium loading of atoms into a double-well optical lattice 
has been studied in Refs. [15-17]. Suppose that the process of such 
a nonequilibrium loading lasts the time $t_{non}$ that is longer than 
the local-equilibrium time $t_{loc}$, but shorter than the relaxation 
time $t_{rel}$ that is necessary for the system for passing to the 
total equilibrium,
$$
t_{loc} \ll t_{non} \ll t_{rel} \; .
$$
In that case, in the interval of time $t_{non} \ll t \ll t_{rel}$, 
the system can be treated as quasiequilibrium, so that the components 
of the itinerant and localized atoms are in equilibrium with each 
other, while the system as a whole has not yet been equilibrated, 
but changes slowly. 

Note that here we consider the case of atoms inside a prescribed 
optical lattice. That is, the considered system is an artificial 
periodic structure, usually, of mesoscopic or nanoscopic size [18]. 
Such a setup is different from the case of a self-organized crystalline 
lattice of a quantum crystal [19-21], in which there can occur jumps 
of particles, connected with self-diffusion [22,23].   

We leave aside the problem of how the desired quasiequilibrium 
structure could be created. Fortunately, optical lattices are highly 
regulated objects, whose parameters can be varied in a wide range [1-5]. 
We assume that such a quasiequilibrium system can be formed. But, since 
the quasiequilibrium has been assumed, this imposes restrictions on 
the system parameters at which the possible coexistence of itinerant 
and localized atoms could be realized. Our aim is to find out what are 
these restrictions and, in particular, what should be the interaction 
parameter $u = U/J z_0$, when the coexistence would be admissible.

\section{Two-Band Model}

To describe the desired coexistence of atoms, we keep in mind a kind 
of a two-band, or two-component, Hubbard Hamiltonian [5], in which one 
band corresponds to delocalized, conducting, atoms, while another band, 
to localized, bound, atoms. The system, as a whole, contains $N_0+N_1$ 
delocalized atoms, where $N_0$ is the number of condensed atoms and 
$N_1$, the number of uncondensed atoms. The number of localized atoms 
is $N_2$. So that the total number of atoms is
\be
\label{1}
 N = N_0 + N_1 + N_2 \; .
\ee
The filling factor is the ratio
\be
\label{2}
\nu \equiv \frac{N}{N_L}
\ee
of the total number of atoms to the number of lattice sites. The 
corresponding atomic fractions
\be
\label{3}
n_0 \equiv \frac{N_0}{N} \; , \qquad
n_1 \equiv \frac{N_1}{N} \; , \qquad
n_2 \equiv \frac{N_2}{N}
\ee
satisfy the normalization condition
\be
\label{4}
 n_0 + n_1 + n_2 = 1 \; ,
\ee
following from Eq. (1). 

The field operator of itinerant atoms, in order to correctly describe the 
Bose-condensed system, is represented by the Bogolubov shifted form [24]
\be
\label{5}
\hat c_j = \eta + c_j \; ,
\ee
with the index $j = 1, 2, \ldots, N_L$ enumerating lattice sites. Here,
$\eta$ is the condensate order parameter, defining the condensate density
$|\eta|^2$, and $c_j$ is an operator of uncondensed atoms. Statistical 
averages for the operators of uncondensed atoms, $c_j$, and for those of 
localized atoms, $b_j$, are such that
\be
\label{6}
\lgl c_j \rgl = \lgl b_j \rgl = 0 \;  .
\ee
 
Thus, the number of itinerant condensed atoms is
\be
\label{7}
N_0 = \sum_j | \eta |^2 = \nu n_0 N_L \;  .
\ee
The number of itinerant uncondensed atoms is
\be
\label{8}
N_1 = \lgl \hat N_1 \rgl = \nu n_1 N_L \;  ,
\ee
with the number operator
\be
\label{9}
\hat N_1 = \sum_j c_j^\dgr c_j \; .
\ee
And the number of localized atoms is
\be
\label{10}
 N_2 = \lgl \hat N_2 \rgl = \nu n_2 N_L \; ,
\ee
with the number operator
\be
\label{11}
 \hat N_2 = \sum_j b_j^\dgr b_j \; .
\ee
 
The energy Hamiltonian has the form of a two-band Hubbard model
\be
\label{12}
\hat H = - J \sum_{\lgl ij \rgl} \hat c_i^\dgr \hat c_j \; + 
\; \frac{U}{2} \sum_j \left ( 
\hat c_j^\dgr \hat c_j^\dgr \hat c_j \hat c_j + 
2\hat c_j^\dgr \hat c_j b_j^\dgr b_j + 
b_j^\dgr b_j^\dgr b_j b_j \right ) \;  ,
\ee
where in the first term the summation is over the nearest neighbors. 
For simplicity, the equal on-site interactions are taken for all atoms. 
Constructing the grand Hamiltonian, we have to take into account the 
given normalization conditions, uniquely defining a representative 
ensemble for the system with broken gauge symmetry [25-27]. Then the 
grand Hamiltonian reads as 
\be
\label{13}
H = \hat H - \mu_0 N_0 - \mu_1 \hat N_1 -
\mu_2 \hat N_2 \;  ,
\ee
in which the Lagrange multipliers $\mu_0, \mu_1$, and $\mu_2$, play 
the role of partial chemical potentials guaranteeing the validity of 
normalizations (7), (8), and (10). The system chemical potential is
\be
\label{14}
\mu = \mu_0 n_0 + \mu_1 n_1 + \mu_2 n_2 \; .
\ee

However, in the considered case, not all these multipliers are 
independent. The restriction comes from the assumption that the 
system is in quasiequilibrium, such that the variation of a 
thermodynamic potential be zero. From this condition, we have 
[5] the relation
$$
\mu_2(n_0 + n_1) = \mu_0 n_0 + \mu_1 n_1  \;  ,
$$  
which gives us the chemical potential
\be
\label{15}
 \mu = \frac{\mu_0 n_0 + \mu_1 n_1}{n_0 + n_1} = \mu_2 \;  .
\ee

Atoms from different bands are assumed to be weakly correlated, such 
that
\be
\label{16}
 c_j^\dgr c_j b_j^\dgr b_j = c_j^\dgr c_j \lgl b_j^\dgr b_j \rgl
+ \lgl c_j^\dgr c_j \rgl b_j^\dgr b_j - \lgl c_j^\dgr c_j \rgl
\lgl b_j^\dgr b_j \rgl \; .
\ee
Then the grand Hamiltonian (13) takes the form
\be
\label{17}
 H = H_{del} + H_{loc} - \nu n_1 n_2 N U \; ,
\ee
where the first term describes delocalized itinerant atoms, the second 
term, localized atoms, while the third term is responsible for the
interactions of atoms from different bands. 

The Hamiltonian of delocalized atoms, substituting there the Bogolubov 
shift (5), becomes the sum
\be
\label{18}
H_{del} = \sum_{n=0}^4 H^{(n)} \;  ,
\ee
with the zero-order term
\be
\label{19}
H^{(0)} = \left ( - z_0 J + \frac{U}{2} \; \nu n_0 -
\mu_0 \right ) n_0 N \;  ,
\ee
first-order term 
\be
\label{20}
 H^{(1)} = 0 \;  , 
\ee
second-order term
\be
\label{21}
H^{(2)} = - J \sum_{\lgl ij \rgl} c_j^\dgr c_j +
(2 \nu n_0 U + \nu n_2 U - \mu_1 ) \sum_j c_j^\dgr c_j
+ \nu n_0 \; \frac{U}{2} \sum_j \left ( c_j^\dgr c_j^\dgr
+ c_j c_j \right ) \; ,
\ee
third-order term
\be
\label{22}
H^{(3)} = \sqrt{\nu n_0 }\; U \sum_j \left ( c_j^\dgr c_j^\dgr c_j +
c_j^\dgr c_j c_j \right ) \; ,
\ee
and the fourth-order term
\be
\label{23}
H^{(4)} = \frac{U}{2} \sum_j c_j^\dgr c_j^\dgr c_j c_j \;.
\ee

The Hamiltonian of localized atoms is
\be
\label{24}
H_{loc} = \sum_j H_j \; ,
\ee
where
\be
\label{25}
H_j = \frac{U}{2} \; b_j^\dgr b_j \left ( b_j^\dgr b_j - 1 
\right ) + [ \nu( n_0 + n_1 ) U - \mu ]\; b_j^\dgr b_j \; .
\ee

\section{Itinerant Atoms}

For the operators of delocalized itinerant atoms, we can invoke the 
Fourier transformation
\be
\label{26}
c_j = \frac{1}{\sqrt{N_L}} \sum_k a_k e^{i\bk \cdot\ba_j} \;  ,
\ee
in which the summation is over the Brillouin zone and $\bf {a}_j$ is a 
lattice vector. For concreteness, we consider in what follows a cubic 
lattice with the lattice distance $a$, for which 
$$
 a^d = \frac{V}{N_L} = \frac{\nu}{\rho} \qquad 
\left ( \rho \equiv \frac{N}{V} \right ) \; ,
$$
with $d$ being the spatial dimensionality.

The second-order term (21) transforms to 
\be
\label{27}
H^{(2)} = \sum_k \left [ - 2J \sum_\al \cos(k_\al a) +
2\nu n_0 U + \nu n_2 U - \mu_1 \right ] a_k^\dgr a_k \;  .
\ee
The third-order term (22) is
\be
\label{28}
H^{(3)} = \frac{U}{N_L} \; \sqrt{N_0}\; \sum_{kp} \left (
a_k^\dgr a_p a_{k+p} + a_{k+p}^\dgr a_p a_k \right ) \;  .
\ee
And the fourth-order term (23) becomes
\be
\label{29}
H^{(4)} = \frac{U}{2N_L} \sum_{kpq} 
a_k^\dgr a_p^\dgr a_{k+q}a_{p-q} \;  .
\ee

The following consideration is in line with the general self-consistent 
approach to Bose systems with broken gauge symmetry, advanced in 
Refs. [25-30], and employed in Refs. [31-35]. We use the definition
\be
\label{30}
\om_k \equiv - 2J \sum_\al \cos(k_\al a) + 
\nu (1 + n_0 + n_1) U - \mu_1  
\ee
and
\be
\label{31}
\Dlt \equiv \nu (n_0 + \sgm) U \;  ,
\ee
where  
$$
n_1 = \frac{1}{N} \sum_k n_k \qquad 
( n_k \equiv \lgl a_k^\dgr a_k \rgl ) \; ,
$$
\be
\label{32}
\sgm = \frac{1}{N} \sum_k \sgm_k \qquad 
( \sgm_k \equiv \lgl a_{-k} a_k \rgl ) \;   .
\ee

Employing the Hartree-Fock-Bogolubov approximation yields
\be
\label{33}
 H_{del} = E_{HFB} + \sum_k \om_k a_k^\dgr a_k +
\frac{1}{2} \sum_k \left ( \Dlt a_k^\dgr a_{-k}^\dgr +
\Dlt^* a_{-k} a_k \right ) \; ,
\ee
where the nonoperator term is
\be
\label{34}
 E_{HFB} \equiv H^{(0)} - N \nu \left ( 2 n_1^2 +
|\sgm|^2 \right ) \frac{U}{2} \; .
\ee
Diagonalizing this Hamiltonian with the help of the Bogolubov 
canonical transformation 
$$
 a_k = u_k \bt_k + v_{-k}^* \bt_{-k}^\dgr \; ,
$$
gives the Bogolubov Hamiltonian 
\be
\label{35}
 H_{del} = E_B + \sum_k \ep_k \bt_k^\dgr \bt_k \; ,
\ee
with the nonoperator term
\be
\label{36}
  E_B = E_{HFB} + \frac{1}{2} \sum_k ( \ep_k - \om_k) 
\ee
and the excitation spectrum
\be
\label{37}
 \ep_k = \sqrt{\om_k^2 - \Dlt^2} \; .
\ee

In equilibrium, the equation of motion for condensate atoms [5] 
reduces to
\be
\label{38}
 \left \lgl \frac{\prt H}{\prt N_0} \right \rgl = 0 \; ,
\ee
which gives the condensate chemical potential
\be
\label{39}
 \mu_0 = - z_0 J + \nu U (n_0 + 2n_1 + \sgm ) \; .
\ee
The condition of condensate existence [5] requires that the spectrum
be gapless, so that 
\be
\label{40}
\lim_{k\ra 0} \ep_k = 0 \qquad( \ep_k \geq 0 ) \; ,
\ee
which is equivalent to the Hugenholtz-Pines relation [36] and yields
\be
\label{41}
 \mu_1 = -z_0 J + \nu U ( 1 + n_1 - \sgm) \; .
\ee

Introducing the notation
\be
\label{42}
e_k \equiv 2J \sum_\al [ 1 - \cos(k_\al a) ] =
4J \sum_\al \sin^2\left ( \frac{k_\al a}{2} \right )
\ee
transforms Eq. (30) into
$$
 \om_k = e_k + \Dlt \; .
$$
The excitation spectrum (37) takes the Bogolubov form
\be
\label{43}
\ep_k = \sqrt{e_k(e_k + 2 \Dlt)} \; .
\ee

The normal and anomalous averages (32) are given by the integrals
\be
\label{44}
n_1 = \frac{1}{\rho} \int_{\cal B} n_k \; 
\frac{d\bk}{(2\pi)^3} \;  , \qquad
\sgm = \frac{1}{\rho} \int_{\cal B} \sgm_k \; 
\frac{d\bk}{(2\pi)^3} \; ,
\ee
in which the integration is over the Brillouin zone and 
\be
\label{45}
 n_k = \frac{\om_k}{2\ep_k}\; {\rm coth}
\left ( \frac{\ep_k}{2T}\right ) - \; \frac{1}{2} \; , \qquad
 \sgm_k = -\;\frac{\Dlt}{2\ep_k}\; {\rm coth}
\left ( \frac{\ep_k}{2T}\right ) \; .
\ee

\section{Localized Atoms}

In the Hamiltonian of localized atoms (24), the site terms (25) can be 
represented as
\be
\label{46}
H_j = \frac{U}{2} \; \hat N_j^2 + \left [ U \left (
\nu n_0 + \nu n_1 - \; \frac{1}{2} \right ) - \mu \right ]
\hat N_j \;  ,
\ee
with the density operator
\be
\label{47}
 \hat N_j \equiv b_j^\dgr b_j \; .
\ee
The average number of localized atoms per site is 
\be
\label{48}
 \lgl \hat N_j \rgl = 
\frac{{\rm Tr}\hat N_j e^{-\bt H_j}}{{\rm Tr}e^{-\bt H_j}} \; .
\ee

Since the eigenvalues of Hamiltonian (46) are
\be
\label{49}
E_n = \frac{U}{2} \; n^2 + \left [ U \left ( \nu n_0 + \nu n_1 - 
\; \frac{1}{2} \right ) - \mu \right ] n \;  ,
\ee
where $n=1,2,\ldots$, the average number of localized atoms (48) is
\be
\label{50}
\nu n_2 = \frac{\sum_{n=0}^\infty n e^{-\bt E_n}}
{\sum_{n=0}^\infty e^{-\bt E_n}} \;   .
\ee

At low temperature, such that $T \ll U$, the maximal contribution into 
the sums over $n$ is given by the term with $n = n_{eff}$, for which 
$E_n$ is minimal, that is, when
\be
\label{51}
\frac{\prt E_n}{\prt n} = 0 \; , \qquad
\frac{\prt^2 E_n}{\prt n^2} > 0 \;  .
\ee
The latter gives 
\be
\label{52}
 n_{eff} = \frac{1}{2} \; - \nu ( n_0 + n_1 ) +
\frac{\mu}{U} \; ,
\ee
under the condition $U>0$. Thus, we get
\be
\label{53}
 \nu n_2 \simeq n_{eff} \qquad ( T \ll U) \; .
\ee
Combining Eqs. (52) and (53) yields
\be
\label{54}
\mu = \left ( \nu -\; \frac{1}{2} \right ) U \qquad 
( T \ll U) \;   .
\ee

According to Eq. (31), we have 
\be
\label{55}
 \sgm + n_0 = \frac{\Dlt}{\nu U} \; .
\ee
Using this in the chemical potentials (39) and (41) results in
$$
\mu_0 = - z_0 J + \Dlt + 2\nu n_1 U \; ,
$$
\be
\label{56}
\mu_1 = - z_0 J - \Dlt + \nu (1 + n_0 + n_1) U \; .
\ee
Equating expressions (15) and (54) gives 
\be
\label{57}
\left ( \nu -\; \frac{1}{2} \right ) U  =
\frac{\mu_0 n_0 + \mu_1 n_1}{n_0 + n_1} \;  .
\ee
 
>From Eqs. (56) and (57), we find
\be
\label{58}
 \frac{n_0}{n_1} = \frac{(1+2\nu n_1)U - 2\Dlt - 2z_0 J}
{(2\nu - 6\nu n_1-1)U - 2\Dlt + 2z_0J} \; .
\ee
The latter, employing the relation
\be
\label{59}
 n_2 = 1 - n_0 - n_1 \; ,
\ee
results in the equation
\be
\label{60}
 n_2 = \frac{U -2z_0J+2\nu U(4n_1-2n_1^2-1)+2\Dlt(1-2n_1)}
{U -2z_0J+2\nu U(3n_1-1)+2\Dlt} \; ,
\ee
defining the fraction of localized atoms $n_2$.

\section{Dimensionless Equations}

It is convenient to pass to dimensionless quantities by introducing
the dimensionless interaction parameter
\be
\label{61}
u \equiv \frac{U}{z_0 J}
\ee
and the dimensionless sound velocity squared
\be
\label{62}
 s^2 \equiv \frac{\Dlt}{z_0 J} \; .
\ee
Also, considering a cubic lattice, we use the dimensionless wave 
vector with the components
\be
\label{63}
 q_\al \equiv \frac{a}{\pi}\; k_\al \qquad (\al=1,2,\ldots,d) \; .
\ee
Measuring all energy quantities in units of $z_0 J$, we have, 
instead of Eq. (30),
\be
\label{64}
\om_k = e_k + s^2 \;  ,
\ee
where
\be
\label{65}
 e_k = \frac{4}{z_0} \sum_\al \sin^2
\left ( \frac{\pi}{2}\; q_\al\right ) \; ,
\ee
and the Bogolubov spectrum (43) becomes
\be
\label{66}
 \ep_k = \sqrt{e_k(e_k+2s^2)} \; .
\ee
 
With the relation $\varrho a^d = \nu$, Eqs. (44) are reduced to
\be
\label{67}
 n_1 = \frac{1}{\nu} \int_0^1 \ldots \int_0^1 n_k \;
dq_1 \ldots dq_d \; , \qquad  
\sgm = \frac{1}{\nu} \int_0^1 \ldots \int_0^1 \sgm_k \;
dq_1 \ldots dq_d \; ,
\ee
where $n_k$ and $\sigma_k$ are given by Eq. (45). In particular, at 
zero temperature, we have
\be
\label{68}
 n_k = \frac{1}{2} \left ( \frac{\om_k}{\ep_k} \; - \; 1 
\right ) \; , \qquad \sgm_k = - \; \frac{s^2}{2\ep_k} \qquad 
(T=0) \; .
\ee

Equation (31) for the sound velocity becomes
\be
\label{69}
 s^2 = ( n_0 + \sgm) \nu u \; .
\ee
And for Eq. (60), we find
\be
\label{70}
n_2 = \frac{2+(2\nu-1)u-4\nu n_1(2-n_1)u+2s^2(2n_1-1)}
{2+(2\nu-1)u-6\nu n_1 u-2s^2} \;  .
\ee

\section{Numerical Results}

We accomplish numerical calculations for a three-dimensional cubic 
lattice ($d = 3$, $z_0 = 6$), with unity filing factor ($\nu = 1$), 
at zero temperature ($T = 0$). Explicitly, we solve the system of 
equations
$$
n_1 = \frac{1}{2} \int_0^1 
\left ( \frac{\om_k}{\ep_k} \; - \; 1 
\right ) dq_1 dq_2 dq_3\; ,
$$
$$
\sgm = -\;\frac{s^2}{2} \int_0^1  
\frac{1}{\ep_k} \; dq_1 dq_2 dq_3 \; ,
$$
$$
\om_k = e_k + s^2\; , \qquad \ep_k=\sqrt{e_k(e_k+2s^2)} \; , 
$$
$$
e_k = \frac{2}{3} \sum_{\al=1}^3 \sin^2
\left ( \frac{\pi}{2}\; q_\al\right ) \; , \qquad 
k_\al \equiv \frac{\pi}{a} \; q_\al \; ,
$$
$$
s^2 = (n_0 + \sgm) u \; , \qquad n_0 = 1 - n_1 - n_2 \; ,
$$
$$
n_2 = \frac{2+u-4un_1(2-n_1)+2s^2(2n_1-1)}
{2+u-6un_1-2s^2} \;  .
$$
The results for the fractions of condensed atoms, $n_0$, uncondensed 
atoms, $n_1$, and localized atoms, $n_2$, are shown in Fig. 1.

When the interaction parameter (61) is smaller than 2.75, there are 
no positive solutions for $n_2$, so that the sole admissible solution 
is $n_2 = 0$, which corresponds to the single-band Hubbard model. 
The mixture can exist only for $u > 2.75$. This gives the lower 
boundary for the possible quasiequilibrium coexistence of 
itinerant Bose-condensed atoms and localized atoms.

\section{Conclusion}

Conditions, under which a quasiequilibrium system of coexisting 
itinerant and localized Bose atoms could be created, are analyzed. 
The consideration is based on a kind of a two-band, or two-component, 
boson Hubbard model [5]. One component corresponds to itinerant atoms 
and is described by the self-consistent Hartree-Fock Bogolubov  
approximation [25-30]. This approximation is known [28-30] to be well
suited for superfluids. But, since it explicitly takes into account 
the global gauge symmetry breaking, it is not suitable for the Mott 
insulating state, where the gauge symmetry breaking is absent [37]. 
Therefore, the localized atoms in the insulating state are 
characterized, in our model, as bound atoms without tunneling between 
lattice sites.

For numerical calculations, we take a three-dimensional cubic lattice. 
Lower-dimensional systems may have no Bose-Einstein condensate and 
require separate consideration [38].

We find that the mixture of itinerant and localized atoms can be 
formed only when the on-site interaction is sufficiently strong, 
such that
$$
u > 2.75 \; .
$$

Recall that here a quasiequilibrium system has been considered. An 
equilibrium system at zero temperature and the unity filling factor, 
as we know, is superfluid below $u_c = 4.9$ and is insulating above 
the latter value, undergoing, at $u_c$, a second-order phase 
transition.  

In the model, we have considered, the itinerant and localized atoms 
are not spatially separated. This distinguishes our case from that 
studied in Ref. [39], where superfluid droplets were separated in 
space, being surrounded by normal, nonsuperfluid, phase. 

We do not consider here the ways of creating such a quasiequilibrium 
mixture. Clearly, this should involve nonequilibrium preparation of 
this state. It is possible to invoke external temporal modulation of 
the system parameters. Different variants of such a modulation of the
parameters of trapped atoms are now discussed in literature [40-45]. 

In order to better understand the physics of the mixture, let us 
give the following picture. Imagine an optical lattice characterized 
by the tunneling frequency $J$ and on-site repulsion $U$. These 
energy parameters are related to the corresponding typical times. 
The energy $z_0 J$ defines the time $t_{osc} = 1/ z_0 J$, during which 
an atom oscillates in a potential well of a lattice site. The smaller 
$z_0 J$, the longer this oscillation time. In the limit of no tunneling, 
this time is infinite. The energy $U$ defines the wandering time 
$t_{wan}=1/U$, during which an atom realizes a jump between the 
neighboring lattice sites. The larger $U$, the shorter this wandering 
time. In the limit of infinite $U$, there are no jumps, all atoms are
completely localized, and the wandering time is zero. The ratio of the 
oscillation time to wandering time is exactly the parameter $u$. The 
condition that $u > 2.75$ implies that
$$
u \equiv \frac{t_{osc}}{t_{wan}} > 2.75  \; .
$$
This means that the oscillation time has to be sufficiently long as
compared to the wandering time. In the other case, atoms could not be 
localized.
   
\vskip 5mm

{\bf Acknowledgement}

\vskip 2mm

One of the authors (V.I.Y.) is grateful for financial support to the
Russian Foundation for Basic Research and another author (A.R.) 
acknowledges financial support from the Volkswagen Foundation.

\newpage

\begin{center}
{\Large {\bf Figure Caption}}
\end{center}

\vskip 3cm

{\bf Fig. 1}. Atomic fractions, as functions of the dimensionless 
interaction parameter $u$, for the condensed atoms (solid line), 
uncondensed atoms (dotted line), and localized atoms (dashed-dotted 
line).

\newpage

\begin{figure}[h]
\centerline{\includegraphics[width=12cm]{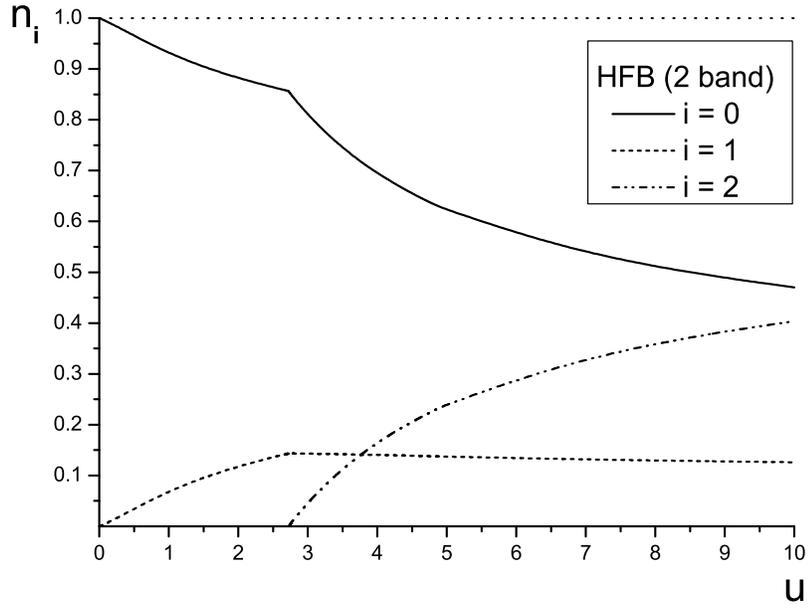}}
\caption{Atomic fractions, as functions of the dimensionless 
interaction parameter $u$, for the condensed atoms (solid line), 
uncondensed atoms (dotted line), and localized atoms (dashed-dotted 
line).}
\label{fig:Figure 1}
\end{figure}

\end{document}